\documentstyle[aps]{revtex}
\input epsf.tex
\begin{document}
\draft
\tightenlines
\newpage

\title{ Search for an annual modulation of dark-matter signals with \\
        a germanium spectrometer at the Sierra Grande Laboratory} 
\author{D. Abriola$^1$, F.T.  Avignone III$^2$, R.L.
Brodzinski$^3$, J.I.  Collar$^4$, \\ D.E.  Di
Gregorio$^1$, H.\ A.\ Farach$^2$, E.  Garc\'{\i}a$^5$, A.O.
Gattone$^1$, C.\ K.\ Gu\'erard$^1$,\\ F.
Hasenbalg$^1$\thanks{Present address:  Laboratory for High Energy
Physics, University of Bern, Sidlerstrasse 5, CH 3012, Bern,
Switzerland.}, H. Huck$^1$, H.S.  Miley$^3$, A.
Morales$^5$,  J.  Morales$^5$, \\ A. Ortiz de
Sol\'orzano$^5$,  J.  Puimed\'on$^5$, J.H.
Reeves$^3$, \\ A. Salinas$^5$, M.L.
Sarsa$^5$\thanks{Present address: Department of Physik E15, 
Technische Universit\"at M\"unchen, James Franck Strasse, 85748 Garching, 
Germany.}, 
J.A.  Villar$^5$ }

\address{ $^1$ Departamento de F\'{\i}sica, Comisi\'on Nacional de Energ\'{\i}a
At\'omica, \\ Av. del Libertador 8250, 1429 Buenos Aires, Argentina.\\ 
$^2$ Department of Physics and Astronomy,
University of South Carolina, \\ Columbia, SC 29208, USA.  \\
$^3$ Pacific Northwest National Laboratory, Richland, Washington 99352,
USA. \\
$^4$ CERN, EP Division, CH-1211 Ginebra, Switzerland. \\
$^5$ Laboratorio de F\'{\i}sica Nuclear y Altas Energ\'{\i}as,
Facultad de Ciencias, \\ Universidad de Zaragoza, 50009 Zaragoza, Spain.
\\}

\maketitle
\begin{abstract}
Results of a search for dark-matter induced annual modulation using
830.5 kg$\cdot$days of data collected at the Sierra Grande underground
laboratory with a germanium detector are presented. The analysis of the
data does not show any indication of seasonal effects.
\end{abstract}

\section{Introduction}

The visible part of our galaxy appears to be embedded in a dark-matter
halo whose radius extends at least 50 kpc beyond the luminous radius.
The halo model and its density value are still open questions, although
a customarily accepted picture features an isothermal, spherically
symmetric, non-rotating halo with a local density of $\rho=[0.3\times
1.5^{0\pm 1}]$ GeV cm$^{-3}$~\cite{sadoulet,jellis}. Recent results
from microlensing data for a flattened halo would favour a higher value
of $\rho$~\cite{turner}.

The MACHO and EROS microlensing observations~\cite{alcock,auburg} of
baryonic dark matter indicate that MACHOs (Massive Astrophysical
Compact Halo Objects) could account for at most half of the dark
halo~\cite{evans}, although the scarce statistics and the model
dependence of its interpretation preclude establishing firm
conclusions~\cite{depaolis}.

It can be concluded, nevertheless, that there is plenty of room for
galactic non-baryonic dark matter~\cite{evans,berezinsky} such as light
neutrinos, WIMPs (Weakly Interacting Massive Particles), or axions. The
slow moving (0.001 c), heavy (GeV--TeV), and neutral WIMPs could be
scattered by the nuclei in a detector producing a detectable signal.
This feature has been used over the past decade to place bounds on the
cross-section and masses of WIMPs by employing low-background germanium
detectors~\cite{ahlen,caldwell,drukier,morales,reusser,beck},
sodium-iodide spectrometers~\cite{fushimi,davies,sarsa,bernabei0}, and
liquid-Xenon scintillators~\cite{belli}.

To go beyond the mere exclusion of candidates obtained so far, and
actually identify the WIMP signal out of the cosmic and environmental
background, one should look for genuine signatures, such as time
modulation in the rate of events at the detector. A possible source of
such modulation~\cite{drukier1} is provided by the orbital motion of
the Earth around the Sun during the solar-system journey through the
galactic halo, which results in a yearly variation of the relative
Earth/halo velocity. The net speed of the Earth with respect to the
halo oscillates between a maximum value (June) and a minimum
(December), and thus the amount of energy that can be deposited by the
WIMPs in the detector, as well as their detection rates, changes
periodically. This kind of modulation has a period of one year, and
searches for it have already been reported in the literature using
semiconductor~\cite{garcia} and scintillator~\cite{sarsa,belli}
detectors with no definitive claim of identification thus far.

The DAMA Collaboration, however, after analyzing 4549.0 kg$\cdot$days of
data collected with nine 9.70-kg NaI(Tl) detectors corresponding to
1185.2 kg$\cdot$days in summer and 3363.8 kg$\cdot$days in winter, has
presented preliminary results~\cite{bernabei1,bernabei2} suggesting
that a yearly modulation effect might be present in their data.  

It is timely, then, that other ongoing experiments report on their
findings, obtained under different conditions with different detectors
and comparable activities. In this paper the analysis and results of a
search for annual modulation are presented using the data collected
with a 1.033-kg germanium detector at the Sierra Grande underground
facility at a depth of 1,000 m.w.e.  This experiment extends over a
period of 1142 days (which is 48 days longer than a full three-year
cycle) with a total effective running exposure of 804 days.  The fact
that three complete oscillations of the putative signal have been
covered increases the statistical reliability of the search, because,
on the one hand, a higher level of structure must be reproduced in
order to account for the expected fluctuations and, on the other, the
chances of an accidental fluctuation are, for the same reason,
considerably reduced. In addition, over long data-taking periods semiconductor
detectors are very stable to temperature and other environmental
changes~\cite{drukier}.  A description of the underground laboratory and
the experimental set-up has been given in Ref.~\cite{abriola}. In
particular the relevant experimental parameters are: an overall energy
resolution of 1.2 keV at 10.37 keV and a long-term energy threshold of
4 keV. A total background spectrum corresponding to an exposure of
830.5 kg$\cdot$day is shown in Fig.~\ref{figure1} where the background
lines at 122.1, 143.5, 840.8, 1124.5, 1173.2 and 1332.5 keV
corresponding to the decay of $^{57}{\rm Co}$, $^{54}{\rm Mn}$,
$^{65}{\rm Zn}$, and $^{60}{\rm Co}$ are clearly recognized.
Calibration of the final spectrum has been performed by using
radioactive sources ($^{182}{\rm Ta}$ and $^{207}{\rm Bi}$) and
low-energy peaks (8.98 and 10.37 keV) clearly identified in the
spectrum.

\section{Data analysis}
\subsection{Modulation significance}

The method of Freese, et al.~\cite{freese} has been customarily used to
look for a modulated component in data.  In that method, a part of the
WIMP signal is modulated with a period $T\simeq 2\pi /\omega$ ($T=365$
days), and the experimental rate, neglecting higher order terms, is
expressed as,
\begin{equation}
S_{tot}(t)=B+S_0+S_m\cos (\omega t) \label{rate}
\end{equation}
where $B$ is the background events, $S_0$ is the
unmodulated part of the WIMP signal, and $t$ is measured from the time
when the maximum speed relative to the halo is achieved 
($\simeq$~the~2$^{\rm nd}$ of June).  A variable, $r$, is defined to act as an
estimator of the significance of the modulation in the signal, by means of
\begin{equation}
r=\frac{X}{\sigma (X)} =\frac{\sum_j 2 \cos (\omega t_j)
S_j}{\sqrt{2\sum_j S_j}}    \label{r}
\end{equation}
with $S_j=S_{tot}(t=t_j)$ the number of events integrated in an
energy interval $\Delta E$ at the $t_j$ day of the acquisition.  The
sum spans the time from the beginning to the end of the experiment.

If a modulation were present in the data, $r$ would depart gradually
from zero with increasing statistics, whereas for an unmodulated case,
the mean value of $r$ would be zero and the variance one.  If the
experiment does not run continuously, however, the expected value of
$r$ will not necessarily be zero, unless one uses only data recorded on
time intervals properly distributed in the cosine period (for instance
in opposite days). If such is not the case, one should first find the
expected value of $r$ -- given the down-time periods that actually
occurred during the experiment -- and compare this value with that
obtained from the real data.

A large number of experiments ($10^4$) were simulated with down-time
periods equal to those of the actual experiment and with the assumption
that $S_m=0$ and that $S_j$ is a Poisson distribution with a mean value
equal to that of the corresponding real data set integrated between
12.5 and 50 keV\footnote{Such interval is not arbitrary but the one
with the highest rate beyond the Zn and Ga X-rays tipically present in
Ge detectors.}.  The frequency distribution of the modulation
significance $r$ statistically obtained in such way  (as well as the
complementary sine projection $s$) are shown in Fig.~\ref{figure2}
using counting rates equal to those of the experiment for that energy
window.

For instance, in the region integrated between 12.5 and 50 keV, the
expected value $r$ is -4.98 and that of $s$ is 4.58 with variances of
1.00 and 0.96, respectively.  A fair criterion to decide whether there
is a significant presence of modulation in the data is to calculate $r$
and $s$ for the real experiment and require that they lie at least
2$\sigma$ away from the mean.  In this situation the hypothesis of no
modulation could be rejected with a confidence level, C.L.
=$\int_{|\langle r \rangle| - 2}^{|\langle r \rangle| + 2}P(r)dr=$
erf$(|\langle r \rangle+2|/\sqrt{2}$) $\ge 95$\% (where $P(r)$ is the
Gaussian probability-density function).  The values obtained from the
data for this particular energy window were $r\ =\ -5.79$ and $s\ =\ 4.04$ 
which do not satisfy the criterion mentioned above.

Because of the small differential rate of the experiment it is
convenient to integrate it in a large energy interval to gain
sensitivity. However, as is well-known, the selected interval should
not include the energy region where the spectra recorded in June and
December cross each other (the so called {\em cross-over} energy where
the June rates start becoming lower than December rates). Integrating
over that region could hide a fluctuation that might actually be
present in the data.  The {\em cross-over} energy depends on the mass of
the WIMP~\cite{lewin,gabutti,sarsa1} and other
parameters~\cite{sarsa1,hase}.  This experiment integrated up to a
large final energy (50 keV) --consistent with the mean energy deposited
by a 10 TeV WIMP-- and ramped the initial energy from 12.5 to 18 keV
thus ensuring sensitivity to fluctuations of WIMPs with masses up to 1
TeV~\cite{hase}.

The modulation-significance variables as a function of the lower limit of 
integration, for several energy intervals in the spectra, are plotted in  
Fig.~\ref{figure3} (crosses) along with the predicted mean values of $r$ 
and $s$ (solid lines) and the 1$\sigma$ contours (dashed).  Values for the 
counting rates are shown in the middle of the figure and range from 
6.3~counts/kg$\cdot$day for the 12.5 to 50 keV energy interval down to 
3.4~counts/kg$\cdot$day for the 18 to 50 keV interval.  The first one 
(6.3~counts/kg.day) corresponds to the distribution shown in 
Fig.~\ref{figure2}.  Clearly, in all cases the significance obtained from 
the data is well consistent with the absence of modulation.

\subsection{Energy-bin analysis}

An alternative and common~\cite{sarsa,belli} way of performing the
analysis is to compute the modulation-significance variables $r$ and
$s$ for a set of small energy bins, compatible with the detector energy
resolution (rather than integrate the signal over broad energy
regions), and look for their distribution. A significant departure from
zero would be an indication of the presence of a modulation. In this
method the question of determining the {\em cross-over} energy is
avoided for all but one of the regions analyzed.

Following Ref.~\cite{bernabei2}, a generalization of the
modulation-significance variables is used, which properly takes
into account, for each energy bin, the correction for down-time
periods. The new variables read:
\begin{equation}
r_0=\frac{\sum_j  [\cos (\omega t_j)-\beta]
S_j}{\sqrt{\sum_j[\cos (\omega t_j)-\beta]^2 S_j}}
\label{r1}
\end{equation}
and
\begin{equation}
s_0=\frac{\sum_j  [\sin (\omega t_j)-\gamma]
S_j}{\sqrt{\sum_j[\sin (\omega t_j)-\gamma]^2 S_j}}
\end{equation}
with $\beta=\frac{1}{N}\sum_j\cos(\omega t_j) =-0.044$ and
$\gamma=\frac{1}{N}\sum_j\sin(\omega t_j) =0.046$, representing the mean
value of the cosine and sine oscillation for the data in
consideration.  $N$ is the actual running time (804 days) out
of the 1142 days of exposure. For later use it is also necessary to
introduce the constant,
\begin{equation}
\alpha=\langle \cos^2 \rangle =\frac{1}{N} \sum_j \cos^2(\omega t_j)
 = 0.531
\end{equation}

In terms of these parameters it is possible to obtain analytical
expressions~\cite{bernabei2} for the modulated part of the signal and its 
dispersion,
\begin{equation}
S_m=\frac{\sum_j [\cos(\omega t_j) - \beta] S_j}{N(\alpha-\beta^2)}
\;\;\;\;\;\;\;\;\;\;\;\;\;\;\;\;
\sigma(S_m)=\frac{\sqrt{\sum_j [\cos(\omega t_j) - \beta]^2S_j}}
{N(\alpha-\beta^2)}, \label{sm}
\end{equation}
as well as for the background plus unmodulated parts (and its
dispersion),
\begin{equation}
b+S_0=\frac{\sum_j [\alpha -\beta\cos(\omega t_j)]S_j}
{N(\alpha-\beta^2)}
\;\;\;\;\;\;\;\;\;\;\;\;\;\;\;\;
\sigma(b+S_0)=\frac{\sqrt{\sum_j [\alpha-\beta\cos(\omega t_j)]^2S_j}}
{N(\alpha-\beta^2)}.
\end{equation}
For an experiment running continuously $\alpha=1/2$ and
$\beta=\gamma=0$. Notice that in this case the dispersion (error bar) of
the modulated amplitude
\begin{equation}
\sigma(S_m)=\sqrt{\frac{2B}{n}} \end{equation}
goes like the inverse square of the number of full periods in which 
the data were collected ($n=N/T$ is the running time in units of $T$, one year.)
Therefore, the sensitivity of the experiment, at constant background rate, 
improves with the square root of the number of periods covered.
Finally, similar expressions to those above may be obtained by substituting 
a sine for the cosine.

A set of fourteen energy bins were selected in the interval 4 to 32 keV
(see Fig.~\ref{figure4}) and the variables $r_0$ and $s_0$ were
calculated for each of them. They are given in Table~\ref{table1}.
Though some of the values obtained for $r_0$ (at 7.0, 15.3 and 21.2
keV) and for $s_0$ (at 5.0, 6.0 and 30.1 keV) are significant, none
depart from zero by more than 2-$\sigma$ implying that this approach
also produced no evidence of modulation at the 97.5\% C.L. Notice that
with three full oscillations under consideration a real effect should
show up in both variables.

The same energy bins were used to extract from the data the modulated
amplitude of the signal, whether present.  The modulated amplitudes (in
counts/keV$\cdot$kg$\cdot$day) as a function of the deposited energy
from Table~\ref{table1} are shown in Fig.~\ref{figure5}.  The squares
correspond to $S_m$ determined from equation (\ref{sm}) and the circles
to substituting a sine for the cosine (the latter shifted half a keV to
the right in the figure for clarity).  The 1-$\sigma$ error bars are
proportional to the square root of the counting rate so they are larger
where the counting rate is larger, i.e. at low energies.  Despite the
fluctuations in the low-energy region, none of the calculated
amplitudes lies more than 1.6-$\sigma$ away from $S_m=0$, and they are
consistent with $S_m=0$. For comparison and to give an idea of the
required sensitivity, the signals expected from WIMPs with masses 30,
40, 50, and 200~GeV are also plotted in the figure, assuming their
cross sections are those from the best exclusion derived so
far~\cite{bernabei0}.  The current sensitivity of the DEMOS experiment can be 
better appreciated by an example: a 50-GeV WIMP modulation would be detected at 
a 2$\sigma$ level for a cross-section $\stackrel{>}{\sim} 2.\times 10^{-4}$ pb.

\subsection{Comparing to DAMA}

To compare the results presented here to those of DAMA we chose to display
the modulated amplitudes extracted from the experiment, $S_m$, as a function
of the energy deposited in each detector and plot them using the same vertical 
scale. Thus, Fig.~\ref{figure6} shows the DAMA/NaI (a) and DEMOS (b) data 
along with the theoretical prediction for a 60 GeV WIMP with a cross-section
$\sigma_{W,nucleon}$=1.0$\times 10^{-5}$ pb, for a halo-abundance density of 
0.3 GeV/cm$^{-3}$. These parameters are representative of the region 
where the analysis of the DAMA data are interpreted in terms of a modulated 
signal. The prediction for DEMOS (Ge) was convoluted with the relative 
efficiency function of Ref.~\cite{ahlen} whereas for DAMA we employed the 
published quenching factors, q(I)=0.09 and q(Na)=0.3. A plot similar to our 
Fig.~\ref{figure6} was already introduced in Ref.~\cite{gerbier} but there the 
normalization of the curve was done relative to the found experimental excess
and is a factor eight larger than our prediction.

As stated above, the data from DEMOS [part b) of the figure] are not 
sensitive enough to exclude the claimed candidate. On the other hand, 
the same analysis applied to the data published by the DAMA 
collaboration does not show an improvement in sensitivity with respect 
to DEMOS. This leads us to  conclude that the alluded candidate 
cannot be either excluded nor identified with current available data and that
more statistics is still necessary if the tip of the neutralino region is to be
probed. 

Regarding the theoretical values of $S_m$ it is worth keeping in mind
that they depend on the subtraction of June minus December rates, so that 
slight changes in the prediction of the rates (form factors, 
energy-smearing or parametrizations of the rates) can lead to sizable 
changes in the prediction of the modulation amplitude.  The inclusion 
of energy smearing due to detector resolution in the calculation of the 
rates can give rise, for the case of NaI, to a factor of two 
decrease in the predicted $S_m$~\cite{gerbier1}. Our predictions in 
Fig.~\ref{figure6} do not include this smearing.

We point out that the plots in Fig.~\ref{figure6} could have been
displayed, alternatively, either matching the error bars of the 
experimental points or normalizing both plots to the peak of the 
predicted signal. In either case the conclusion does not vary in 
that the putative candidate could be there since the sensitivity of 
both experiments is, at the present time, not enough to either exclude 
nor prove it. 

\section{Summary and conclusions}

Data collected during three years with a germanium spectrometer at the
Sierra Grande underground laboratory have been analyzed for distinctive
features of annual modulation of the signal induced by WIMP dark matter
candidates. The main motivation for this analysis was the recent
suggestion~\cite{bernabei1,bernabei2} that a yearly modulation signal
could not be rejected at the 90\% confidence level when analyzing data
obtained with a high-mass low-background scintillator detector. Two
different analyses of the data were performed. First, the statistical
distribution of modulation-significance variables (expected from an
experiment running under the conditions of Sierra Grande) was compared
with the same variables obtained from the data.  Second, the data were
analyzed in energy bins as an independent check of the first result
and  to allow for the possibility of a crossover in the expected
signal.  In both cases no statistically significant deviation from the
null result was found, which could support the hypothesis that the data
contain a modulated component. Finally, a plot was presented to be able
to compare our results to those of the DAMA collaboration.

\acknowledgements{Fruitful discussions with  G. Gerbier are gratefully
acknowledged. The members of CNEA wish to thank CONICET (Argentina)
for financial support. The members of UZ wish to thank CICYT (Spain)
for financial support. DDG, FH and HH acknowledge partial support from
Universidad Nacional de Gral. San Mart\'{\i}n.}

\newpage

\begin{table}
\caption{Modulation-significance variables $r_0$ and $s_0$ calculated
for the fourteen energy bins in which the data were divided (from 4 to
32 keV). Also showm are the modulated amplitudes and their dispersions
for the cosine- (columns 4 and 5) and the sine- (columns 6 and 7)
modulation analysis. The listed energies correspond to the lower end of
the interval.}

\begin{tabular}{rrrrrrr} 
E(keV) & $r_0$   &    $s_0$     &     $S_m$(cos)   &         d$S_m$(cos)   &    
$S_m$(sin)    &        d$S_m$(sin) \\ \hline
 4.0  & -0.41 & -1.05  & -0.032 &  0.080 & -0.090 & 0.085 \\
 5.0  & -1.02 & -1.66  & -0.081 &  0.079 & -0.143 & 0.087 \\
 6.0  & -0.20 & -1.35  & -0.016 &  0.078 & -0.113 & 0.084 \\
 7.0  &  1.76 & -0.27  &  0.121 &  0.069 & -0.021 & 0.076 \\
14.2  & -0.77 & -0.92  & -0.013 &  0.017 & -0.017 & 0.018 \\
15.3  &  1.75 & -0.16  &  0.019 &  0.011 & -0.002 & 0.012 \\
17.7  &  0.49 &  0.01  &  0.006 &  0.013 &  0.001 & 0.001 \\
19.4  & -0.55 &  0.86  & -0.007 &  0.012 &  0.011 & 0.013 \\
21.2  & -1.99 &  0.76  & -0.023 &  0.012 &  0.009 & 0.012 \\
23.0  & -0.98 &  0.68  & -0.011 &  0.011 &  0.008 & 0.011 \\
24.7  & -0.30 &  0.83  & -0.003 &  0.010 &  0.009 & 0.011 \\
26.5  & -0.42 & -0.37  & -0.004 &  0.010 & -0.004 & 0.010 \\
28.3  &  0.39 & -0.36  &  0.004 &  0.009 & -0.003 & 0.010 \\
30.1  &  1.06 & -1.66  &  0.009 &  0.009 & -0.015 & 0.009
\end{tabular}
\label{table1}
\end{table}

\newpage

\begin{figure}
\caption{Total background spectrum corresponding to an exposure of 830.5 
kg$\cdot$day.
 \label{figure1}}
\end{figure}

\begin{figure}
\caption{Frequency plot of the modulation-significance
variables $r$ and $s$. Each curve represents the
calculated values of the variable for 10,000 simulated
experiments running for 1142 days with down-time periods
and background similar to the actual experiment in Sierra Grande.
\label{figure2}}\end{figure}

\begin{figure}
\caption{Predicted (solid lines) and calculated (crosses) values of the
modulation-significance variables $s$ (upper) and $r$ (lower) as a
function of the low-energy limit of integration, $E_i$. The dashed
lines correspond to the 1-$\sigma$ contours; count rates of each data
set are also shown.\label{figure3}}
\end{figure}

\begin{figure}
\caption{Selected energy bins for the analysis of section II.B. The gap
between 7 and 14~keV corresponds to Zn and Ga X-rays and was not
considered in the analysis. \label{figure4}}
\end{figure}

\begin{figure}
\caption{Modulated amplitude, ($S_m$), extracted from the data using
Eq.~(\protect\ref{sm}), as a function of the deposited energy. The
error bars are 1-$\sigma$. The curves correspond to the signal expected
from a spin-independent WIMP with a mass of 30 (solid), 40 (dash), 50
(dot) and 200 GeV (dot-dash) and a cross section corresponding to the
exclusion plot $\sigma(m_\chi)$ of~ref.~\protect\cite{bernabei0}.
\label{figure5}}
\end{figure}

\begin{figure}
\caption{Modulated amplitudes $S_m$ [Eq.~(\protect\ref{sm}) in the text]
as a function of the deposited energy obtained from the data of the 
DAMA/NaI (a) and DEMOS (b) collaborations. In both plots the solid line 
corresponds to the signal expected from a 60 GeV WIMP scattering off each  
detector with $\sigma_{W,nucleon}$=1.0$\times 10^{-5}$ pb. For the NaI
case 
the energy resolution (which was not taken into account in the
calculation)
would decrease the amplitude of the curve by a factor 2 (see text). The circles in 
part b) of the figure correspond to the substitution of a sine for the 
cosine used to obtain $S_m$ and are shifted 0.5 keV to the right for clarity.
\label{figure6}}
\end{figure}

\newpage
\thispagestyle{headings}
\markright{\huge{\bf FIG. 1}}
\epsffile{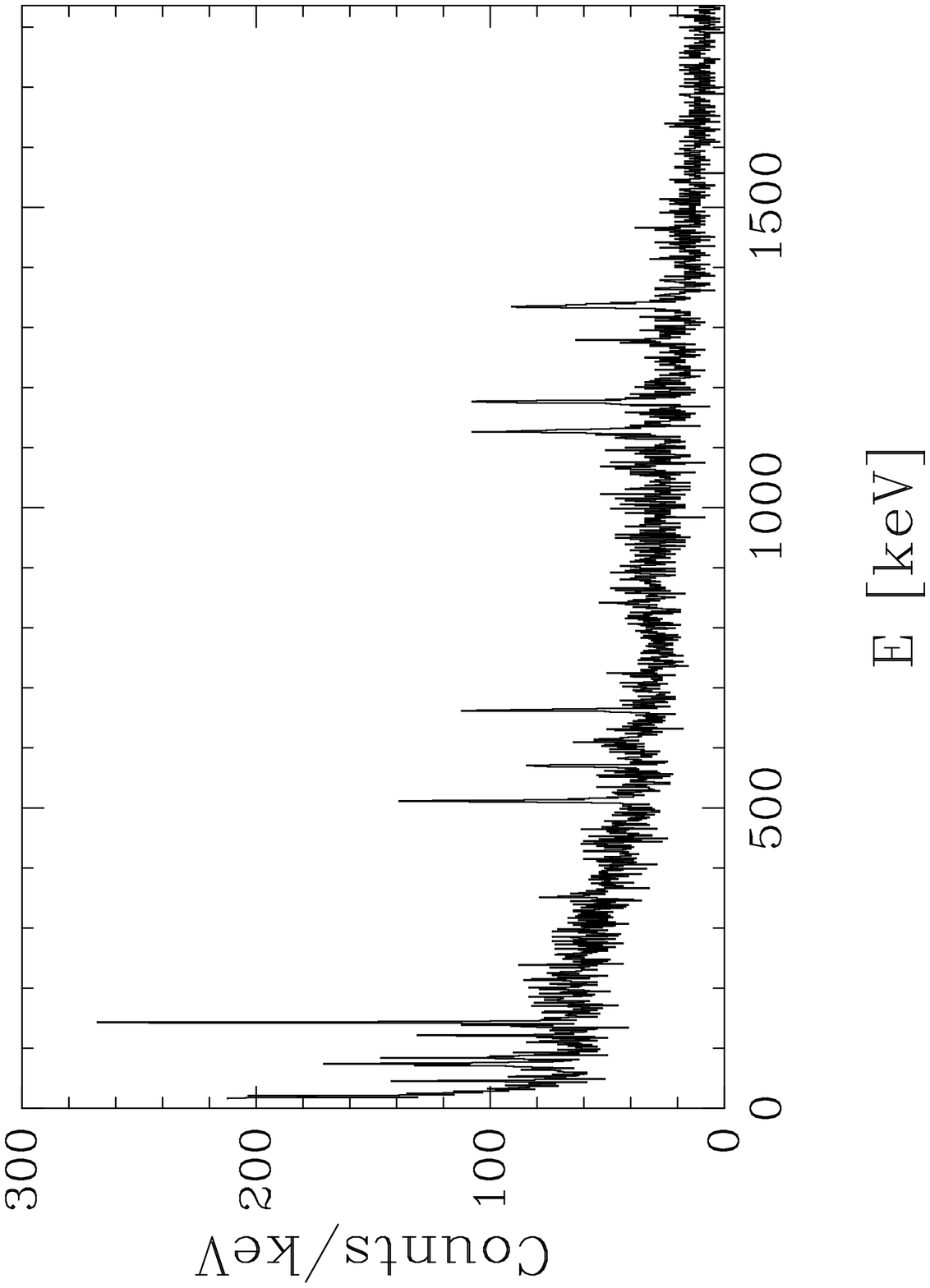}

\newpage
\thispagestyle{headings}
\markright{\huge{\bf FIG. 2}}
\epsffile{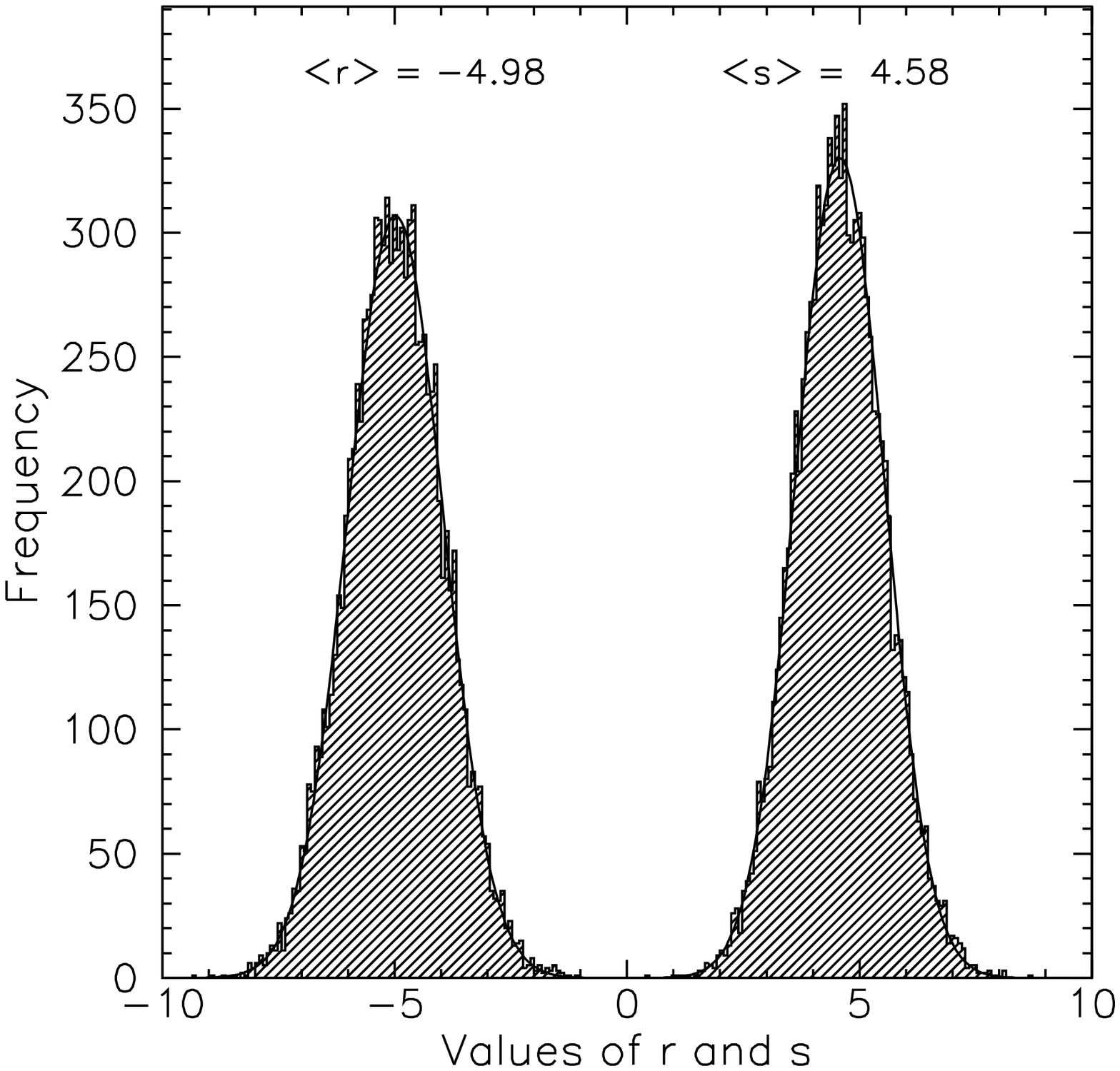}

\newpage
\thispagestyle{headings}
\markright{\huge{\bf FIG. 3}}
\epsffile{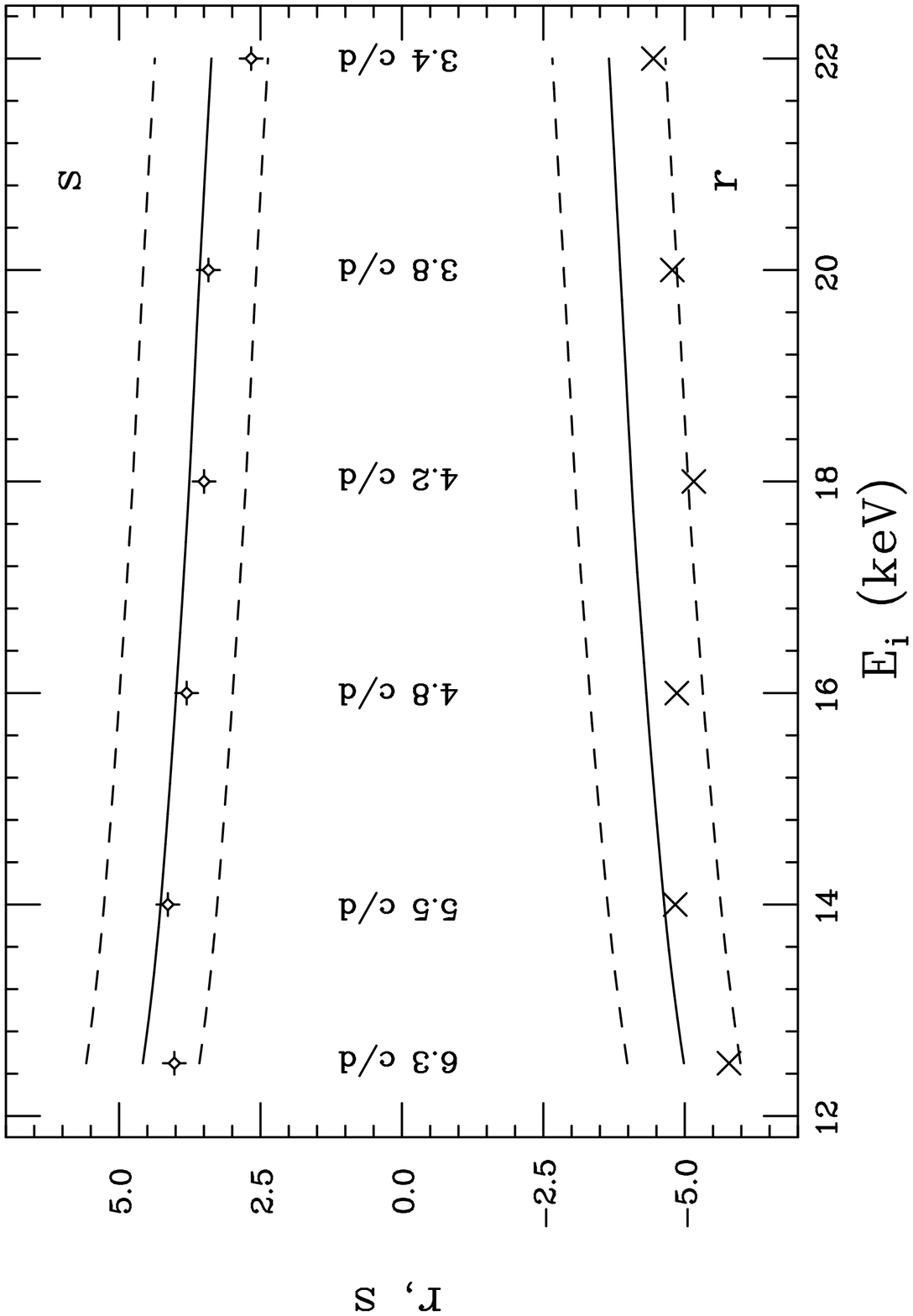}

\newpage
\thispagestyle{headings}
\markright{\huge{\bf FIG. 4}}
\epsffile{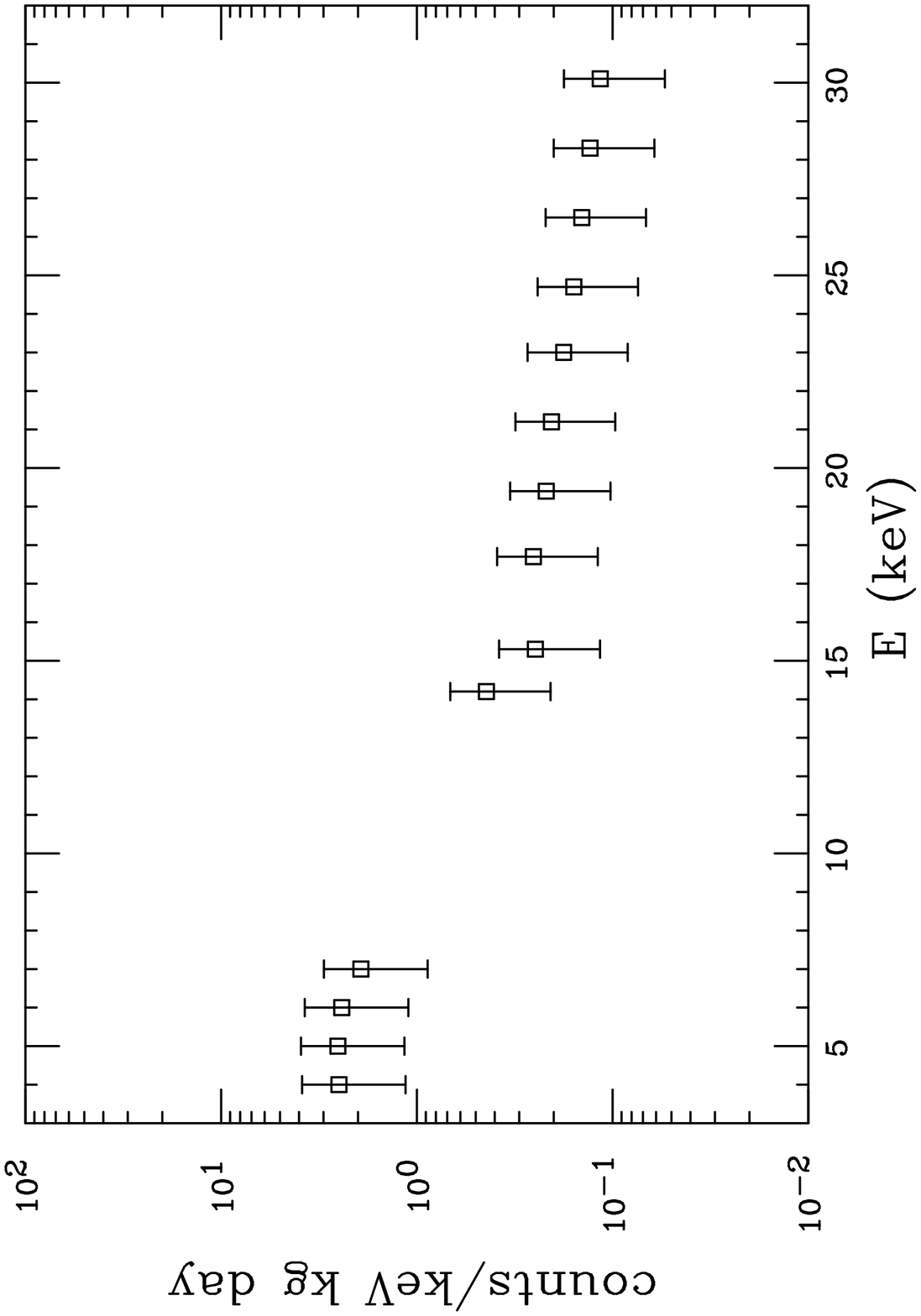}

\newpage
\thispagestyle{empty}
\thispagestyle{headings}
\markright{\huge{\bf FIG. 5}}
\epsffile{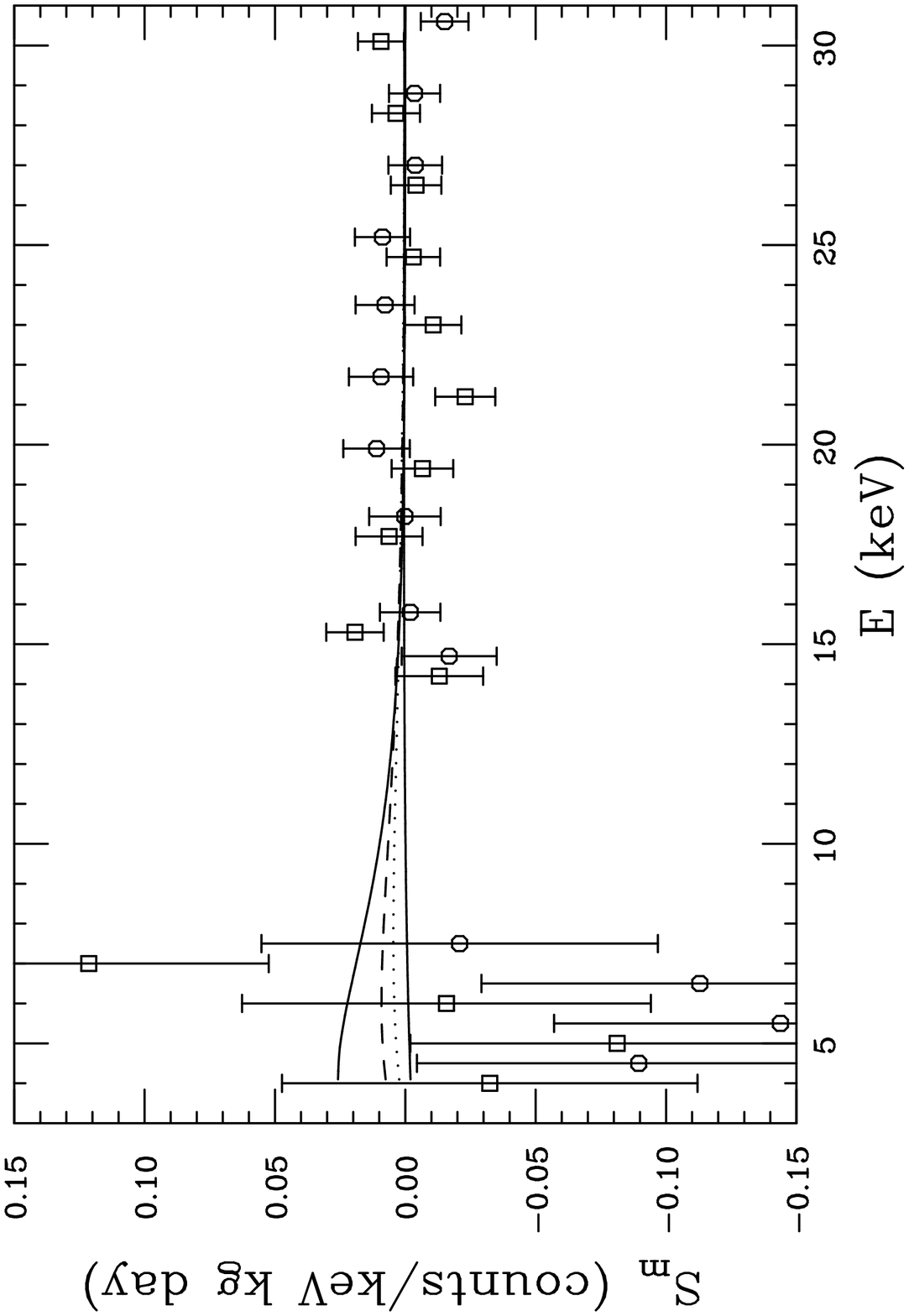}

\newpage
\thispagestyle{empty}
\thispagestyle{headings}
\markright{\huge{\bf FIG. 6.a}}
\epsffile{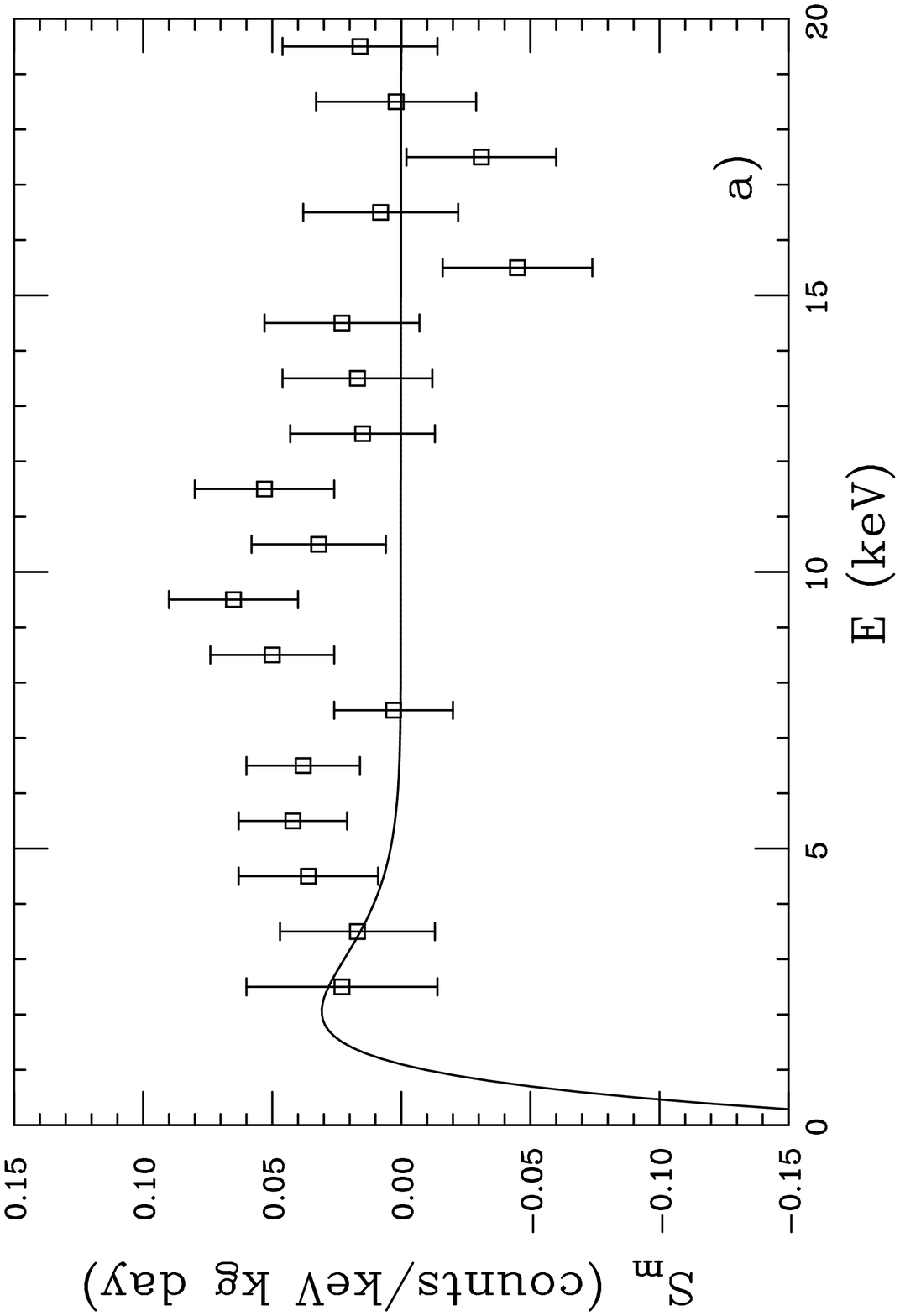}

\newpage
\thispagestyle{empty}
\thispagestyle{headings}
\markright{\huge{\bf FIG. 6.b}}
\epsffile{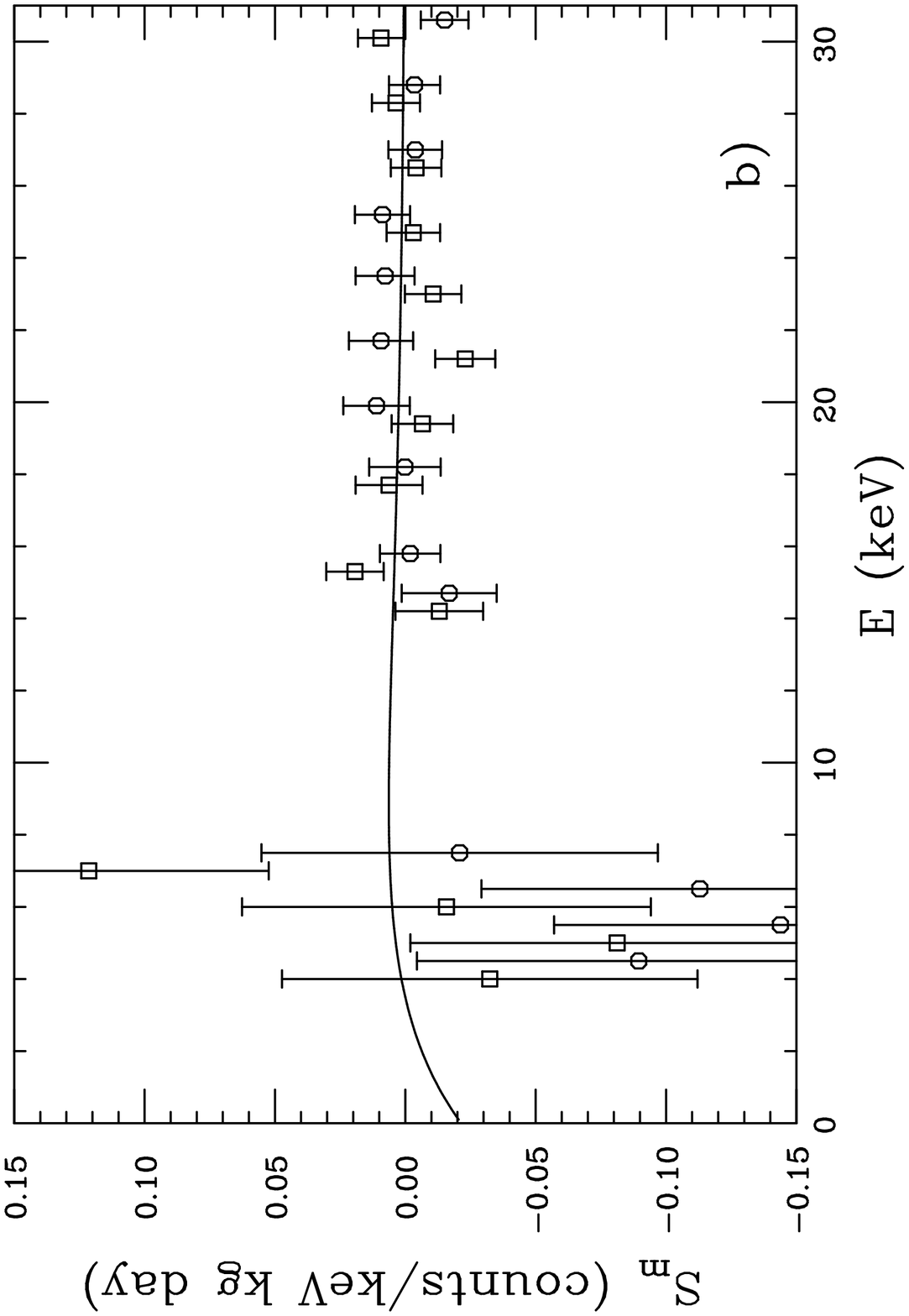}


\begin{references}

\bibitem{sadoulet} J.\ R. Primack, D.\ Seckel, and B.\ Sadoulet, Ann. Rev. Nucl. 
Part. Sci. {\bf 38} (1995) 751.

\bibitem{jellis} J.\ Ellis, Nucl. Phys. B (Proc. Suppl.) {\bf 48} (1996) 522 and 
references therein.

\bibitem{turner} E. I. Gates, G.\ Gyuk, and M.\ S.\ Turner, Phys. Rev. Lett. 
{\bf 74} (1995) 3724; Ap. J. {\bf 449} (1995) L123.

\bibitem{alcock} C.\ Alcock, et al., Nature {\bf 365} (1993) 621; Phys. Rev. 
Lett. {\bf 74} (1995) 2867; Ap. J. {\bf 461} (1996) 89; preprint 
astro-ph/9606165.

\bibitem{auburg} E.\ Auburg, et al., Nature {\bf 365} (1993) 623. C.\ Renault, 
et al., Astron. Astrophys. {\bf 324} (1997) L69.

\bibitem{evans}N.\ W. Evans, G.\ Gyuk, M.\ S.\ Turner, and J.\
Binney, submitted to {\em Nature} (1997).

\bibitem{depaolis} F.\ De Paolis, G.\ Ingrosso, and Ph. Jetzer, Ap. J. {\bf 470} 
(1996) 493.

\bibitem{berezinsky} See for instance V. Berezinsky, et al., Nucl. Phys. B 
(Proc. Suppl.) {\bf 48} (1996) 22 and references therein.

\bibitem{ahlen} S.\ P.\ Ahlen, {\em et al.}, Phys. Lett. B {\bf 195}
(1987) 603.

\bibitem{caldwell} D.\ O.\ Caldwell, et al., Phys. Rev. Lett. {\bf 61} (1988) 
510. D.\ O.\ Caldwell, Nucl. Phys. B (Proc. Suppl.) {\bf 38} (1995) 394. 

\bibitem{drukier} A.\ K.\ Drukier, et al., Nucl. Phys. B (Proc. Suppl.) {\bf 
28A} (1992) 293. 

\bibitem{morales} J.\ Morales, et al., Nucl. Instrum. Methods Phys. Res. A {\bf 
321} (1992) 410; E.\ Garc\'{\i}a, et al., Phys. Rev. D {\bf 51} (1995) 1460.

\bibitem{reusser} D.\ Reusser, et al., Phys. Lett. B {\bf 255} (1991) 143.

\bibitem{beck} M.\ Beck, {\em et al.}, Phys. Lett. B {\bf 336} (1994)
141.

\bibitem{fushimi} K.\ Fushimi, et al., Phys. Rev. C {\bf 47} (1993) R425. 
H.\ Ejiri, et al., Phys. Lett. B {\bf 317} (1993) 14.

\bibitem{davies} G.\ J.\ Davies, et al., Phys. Lett. B {\bf 322} (1994) 159. 
N.\ G.\ C.\ Spooner and P.\ F.\ Smith, Phys. Lett. B 
{\bf 314} (1993) 430. N.\ G.\ C.\ Spooner, et al., Phys. Lett. B {\bf 321} 
(1994) 156.

\bibitem{sarsa} M.\ L.\ Sarsa, et al., Phys. Lett. B {\bf 386} (1996) 458 and
Phys. Rev. D {\bf 56} (1997) 1856.

\bibitem{bernabei0} R. Bernabei, {\em et al.}, Phys. Lett. B {\bf 389}
(1996) 757.

\bibitem{belli} P.\ Belli, et al., Nuovo Cim. C {\bf 19} (1996) 537; 
Nucl. Phys. B (Proc. Suppl.) {\bf 48} (1996) 62.

\bibitem{drukier1} A. K. Drukier, K. Freese, and D. N. Spergel, Phys. Rev. D
{\bf 33}, 3495 (1986).

\bibitem{garcia}E.\ Garc\'{\i}a, et al., in {\em The Dark Side
of the Universe: Experimental Efforts and Theoretical Framework},
Proceedings of the International Workshop, Rome, Italy, 1993, edited
by R. Bernabei and C. Tao (World Scientific, Singapore, 1994), p. 216.

\bibitem{bernabei1} R. Bernabei, et al., in TAUP97: Topics in Astroparticle and 
Underground Physics. Proceedings of the Fifth International Workshop, Gran 
Sasso, 
Italy, 1997, eds. A.~Bottino, A.~de Credico, and P. Monacelli, Nucl. Phys. B 
(Proc. Suppl.) {\bf 70} (1998) 79.

\bibitem{bernabei2} R. Bernabei, et al., ROM2F/97/33, Aug. 1997 and Phys. Lett. 
B {\bf 424} (1998) 195.

\bibitem{abriola}D.\ Abriola, et al., Astropart. Phys. {\bf 6},
63 (1996).

\bibitem{freese} K.\ Freese, J.\ Frieman, and A.\ Gould, Phys. Rev. D
{\bf 37} (1988) 3388.

\bibitem{lewin} J.\ D.\ Lewin, P.\ F.\ Smith, Astropart. Phys. {\bf
6}, (1996) 87.

\bibitem{gabutti} A.\ Gabutti and M. Schmieman, Phys. Lett. B {\bf 308} (1993) 
411.

\bibitem{sarsa1} M.\ L.\ Sarsa, PhD. Dissertation, Publ. Univ. Zaragoza, 
November (1995). Dep\'osito Legal No. Z-3857-95.

\bibitem{hase} F. Hasenbalg, submitted to Astropart. Phys. (1997).

\bibitem{gerbier}G.\ Gerbier, J.\ Mallet, L.\ Mosca, and C.\ Tao, {\em
astro-ph/9710181}.

\bibitem{gerbier1}G.\ Gerbier, private communication.

\end{references}
\end{document}